\documentclass{aa}  
\usepackage{graphicx}
\usepackage{natbib}
\usepackage{amssymb}

\usepackage{txfonts}
\begin{document}

\title{X-rays from RU Lupi - Accretion and winds in CTTS}

   \author{J. Robrade
          \and J.H.M.M. Schmitt
          }
   \institute{Universit\"at Hamburg, Hamburger Sternwarte, Gojenbergsweg 112, D-21029 Hamburg, Germany\\
       \email{jrobrade@hs.uni-hamburg.de} 
  }

\date{Received 13 April 2007 / Accepted 18 June 2007}
  \abstract
 {Low-mass stars are known to exhibit strong X-ray emission during their early evolutionary stages.
This applies also to classical T Tauri stars (CTTS), whose X-ray emission differs from that of main-sequence stars in a number of aspects.}
 {We study the specific case of RU~Lup, a well known accreting and wind-driving CTTS.
In comparison with other bright CTTS we study possible signatures of accretion and winds in their X-ray emission.}
  {Using three XMM-Newton observations of RU~Lup, we investigate
its X-ray properties and their generating mechanisms.
High resolution X-ray spectra of RU~Lup and other CTTS are compared to main-sequence stars. We examine the presence of a cool plasma excess and enhanced plasma density
in relation to X-rays from accretion shocks and investigate anomalous strong X-ray absorption and its connection to  winds or circumstellar material.}
   {We find three distinguishable levels of activity among the observations of RU~Lup.
While no large flares are present, this variability is clearly of magnetic origin due to the corresponding plasma  temperatures of around 30\,MK;
in contrast the cool plasma component at 2\,--\,3\,MK is quite stable over a month, resulting in average plasma temperature from 35\,MK down to 10\,MK.
Density analysis with the \ion{O}{vii} triplet indicates high densities in the cool plasma, suggesting accretion shocks to be a significant contributor to the soft X-ray emission.
No strong overall metal depletion is observed, with Ne being more abundant than Fe, that is at solar value, and especially O. Excess emission at 6.4\,keV during the more active phase
suggest the presence of iron fluorescence.
Additionally RU~Lup exhibits an extraordinary strong X-ray absorption, incompatible with estimates obtained at optical and UV wavelengths.
Comparing spectra from a sample of main-sequence stars with those of accreting stars we find an
excess of cool plasma as evidenced by lower \ion{O}{viii}/\ion{O}{vii} line ratios in all accreting stars. High density plasma appears to be
only present in low-mass CTTS, while accreting stars with intermediate masses ($\gtrsim 2 M_{\odot}$) have lower densities.}
   {In all investigated CTTS the characteristics
of the cooler X-ray emitting plasma are influenced by the accretion process. We suspect different accretion rates and amount of funneling, possibly linked to stellar mass and radius,
to be mainly responsible for the different properties of their cool plasma component.
The exceptional X-ray absorption in RU~Lup and other CTTS is probably related to the accretion flows and an optically transparent wind emanating from the star or the disk.
}


   \maketitle
%

\section{Introduction}

T~Tauri stars as a class are very young low-mass stars. The members of the subclass of so-called classical T~Tauri stars (CTTS) are 
still accreting matter from a surrounding circumstellar disk. CTTS are thought to evolve first into weak-line T~Tauri stars (WTTS), where they
become virtually disk-less and show no longer signs of significant accretion, and eventually into solar-like star on the main sequence. 
Ongoing accretion in a CTTS is evidenced by an emission lines spectrum and specifically a large H$\alpha$ equivalent width (EW~$>$10~\AA).
Further, the strong observed infrared excess indicates the presence of a disk. In contrast, WTTS have much weaker H$\alpha$ emission and little or no IR-excess.
The increasing dominance of the underlying
continuum emission over the optical stellar absorption line spectrum gives rise to another sub-classification into moderate, veiled and extreme TTS.
However, these strict subdivisions are somewhat arbitrary and blurred since for example H$\alpha$ emission is highly variable 
and differs also within the CTTS class by more than an order of magnitude. 
A detailed account of the pre-main sequence (PMS) stellar evolution of low-mass stars is given by \citet{fei99}. 

In the commonly accepted magnetospheric accretion model for CTTS \citep{cal98} material is assumed to be accreted from the stellar disk onto the star at
almost free-fall velocity along magnetic field lines that
disrupt the accretion disk in the vicinity of the corotation radius. Upon impact a strong shock is formed near the stellar surface and
the funneling of the accreted matter by the magnetic field leads to "accretion spots" with filling factors at the percent level with respect
to the stellar surface; therefore only a small fraction of the stellar surface is covered by the accretion spots.
This accretion shock plasma is expected to reach temperatures of up to a few MK
and thus to lose a large fraction of its energy at UV and soft X-ray wavelengths.
The accreted plasma does not
necessarily have to be at high density. However, if one wants to
achieve the inferred small filling factors at the percent level and the otherwise determined mass accretion rates of $10^{-8}-10^{-7}M_{\odot}$\,yr$^{-1}$,
the infalling gas must then be at rather high density (n~$> 10^{11}$\,cm$^{-3}$). The accretion process is
accompanied by outflows from the star and possibly also from the disk, that probably also play an important role in the transport of angular momentum.
Different mass accretion rates and filling factors, different stellar properties like mass, rotation and activity 
and different viewing angles naturally lead to the observed variety of the different CTTS phenomena.

X-ray emission from T~Tauri stars has already been detected with the
{\it Einstein} and ROSAT observatories, and both types of TTS are found to be copious and variable X-ray emitters.  
The origin of the observed X-ray emission is currently subject of some debate.  
Since all main-sequence "cool stars", i.e. stars with outer convection zones, are surrounded by  X-ray emitting coronae \citep{schmitt04}, 
some kind of magnetic activity is expected also for their PMS counterparts with their deep outer convection zones. Indeed,
large sample studies like COUP ({\it Chandra} Orion Ultradeep Project) showed that most of their observed X-ray emission is of coronal origin \citep{pre05}.
Further, magnetic fields, connecting star and disk, have been invoked to explain the huge flares observed in CTTS \citep{fav05}. 

On the other hand, the presence of accretion streams and outflows opens up additional possibilities for the
generation of X-rays. Collimated winds have been observed as X-ray jets in Herbig-Haro objects \citep{prav01,fav06}
and in CTTS like DG~Tau \citep{gue05}. Accretion shocks are expected to generate plasma at a significantly higher density than stellar coronal plasma
and their soft X-ray emission can be traced by density sensitive line ratios,
e.g. between forbidden and intercombination lines in the He-like triplets of \ion{O}{vii} and \ion{Ne}{ix}. 
Anomalously low ratios of these lines have been
observed for several CTTS, e.g. TW~Hya \citep{kas02,ste04}, BP~Tau \citep{schmitt05}, CR~Cha \citep{rob06},
V4046~Sgr \citep{guenther06} or MP~Mus \citep{arg07} and the inferred high plasma densities
suggest the presence of X-ray emission from shocks as postulated in the magnetospheric accretion scenario.

However, not all accreting stars seem to show high density plasma. 
The prototypical, but not necessarily typical TTS T~Tau itself is an example where density diagnostics of cool plasma 
as seen in \ion{O}{vii} indicate a lower density that is more comparable to active main sequence stars \citep{gue07}. 
Further, accretion is only capable of producing plasma with temperatures of up to a few MK \citep[e.g.][]{guenther07}, and therefore
contributes nearly exclusively to the low energy X-ray emission.  However, in all CTTS - even in the accretion dominated, "soft" TW~Hya -
high temperature plasma is observed, which together with the
strong flaring requires magnetic activity to be also present in these objects. 
A comparative study of several bright CTTS indicated the presence of X-rays from both accretion and magnetic activity, but with different
respective contributions in the individual objects \citep{rob06}. 
Altogether, magnetic activity is the dominant source of the X-ray emission in most CTTS at least at higher energies, however accretion can significantly contribute to the soft X-ray emission.

RU Lup is a late K-type CTTS with a high accretion rate and an extreme optical veiling. It has been extensively studied especially at optical and UV 
wavelengths and a review of its basic stellar properties is given by \citet{lam96} and \citet{stem02}. \citet{lam96}
present a magnetospheric accretion model for RU~Lup based on multi-frequency monitoring and high-resolution optical data.
While there was some debate about the distance of RU~Lup (140\,--\,250\,pc), recent estimates indicate a value of $\sim$~140\,pc \citep{hug93}, which we adopt for this paper.
Already \citet{lam96} pointed out that their derived reddening sets an upper limit to RU~Lup's distance of $\sim$ 150\,pc, when attributed only to interstellar material, 
also indicating very little circumstellar absorption.
RU~Lup is highly variable and shows strong accretion signatures with a large H$\alpha$ EW of 140 \AA\ up to 216~\AA\ \citep{app83}.
Virtually no photospheric absorption lines are detected in its spectrum due to substantial optical veiling, on the
contrary, the optical spectrum shows many emission lines.  
The line profiles are also strongly variable and asymmetric, suggesting both infalls and outflows with velocities up to 300\,--\,400~km/s.
Specifically, \citet{lam00} attributes the origin of the observed lines profiles in the HST UV spectra to accretion shocks and stellar winds.
From optical high resolution spectra \citet{stem02} determined mass and radius of RU~Lup to
$M \sim 0.8 M_{\odot}$ and $R \sim 1.7 R_{\odot}$.  They further
derived a projected rotational velocity of  $v sin(i)=9$~km/s and detected variability on short time scales of 1h. 
Together with various proposed periods (0.8--3.7 d) this suggests an inclination between 3\degr and 16\degr, i.e. a system that is
viewed nearly pole-on with rather high intrinsic rotation. Finally, they derived an
interstellar absorption column towards RU~Lup with a corresponding N(H)=$6.3\times 10^{19}$\,cm$^{-2}$.
Combining literature data with results derived from FUV data obtained with FUSE and HST/STIS, \citet{her05} estimated 
RU~Lup to be an 0.6\,--\,0.7~$M_{\odot}$ star with an age of 2\,--\,3~Myr and
a mass accretion rate of $\sim 5 \times 10^{-8} M_{\odot}$.  
Their derived hydrogen column density of log N(H)$= 20.1\pm0.2$\,cm$^{-2}$,
corresponding to an extinction of A$_{\rm V}\sim 0.07$, is again quite low.
Combining and summarising these 
findings we conclude that RU~Lup is probably an optically rather mildly absorbed system viewed almost pole-on.

While the {\it Einstein} IPC failed to detect X-ray emission from RU~Lup with an upper flux limit of $1.2 \times 10^{-13}$\,erg\,cm$^{-2}$ s$^{-1}$
in the 0.5--4.0\,keV band \citep{gahm80},
X-ray emission from RU~Lup was detected 1993 in a pointed 15.7\,ks ROSAT PSPC (0.1--2.4\,keV) observation 
at a rate of $9\times 10^{-3}$\,cts/s (2RXP catalog). 
Using the energy conversion factors given by \citet{neuh95}, this corresponds to an X-ray flux 
of $\sim 1 \times 10^{-13}$\,erg\,cm$^{-2}$\,s$^{-1}$ and an X-ray luminosity of log~$L_{\rm X} = 29.4$\,erg\,s$^{-1}$ 
for a distance of 140~pc.

In this paper we present three new {\it XMM-Newton} observations of RU~Lup with a total exposure time of 84\,ks,
performed over the course of a month in 2005. Our paper is structured as follows.
In Sect.\,\ref{obsana} the X-ray observations and the data analysis are described,
in Sect.\,\ref{results} we present our results derived from the {\it XMM-Newton} observations of
RU~Lup subdivided into different physical
topics. Finally, in Sect.\,\ref{ctts} we 
compare RU~Lup to results from X-ray diagnostics of several CTTS and specifically
discuss accretion and wind scenarios in the context of their X-ray emission in general that is followed by our conclusions in Sect.\,\ref{sum}.

\section{Observations and data analysis}
\label{obsana}

RU~Lup was observed by {\it XMM-Newton} three times in the course of a month in August/September
2005 for roughly 30\,ks each; the detailed observation times are listed in Table\,\ref{log}.
Data were taken with all X-ray detectors,
respectively the EPIC (European Photon Imaging Camera), consisting of the MOS and PN detectors
and the RGS (Reflection Grating Spectrometer) as well as the optical monitor (OM) .
The EPIC instruments were operated in full frame mode with the medium filter, while the
OM was operated with the UVW1 filter, which is sensitive in the waveband between
2000\,--\,4000~\AA\ with an effective wavelength of 2910~\AA . 
The OM exposures typically lasted around 4\,ks and allow an accurate determination of RU~Lup's respective UV brightness.
During each X-ray observation six exposures of the central field containing RU~Lup were taken.

\begin{table} [ht!]
\begin{center}
\caption{\label{log} Observing log of the {\em XMM-Newton} RU~Lup exposures.}
\begin{tabular}{rrrr}
\hline
\multicolumn{2}{l}{Observation}   & MOS (filt.)& OM \\
Date &Time    & Dur.(ks)& Exp.(no.)  \\\hline
August 08 2005 & 04:19 -- 12:54 & 30(28) & 6\\
August 17 2005& 09:17 -- 17:02& 27(15) & 6\\
Sept. 06/07 2005 & 18:30 -- 02:16 & 27(22) & 6\\
\hline
\end{tabular}
\end{center}
\end{table}

All data analysis was carried out with the {\it XMM-Newton} Science Analysis System (SAS) version~7.0 \citep{sas}.
Standard selection criteria were applied to the data and
periods with enhanced background due to proton flares were discarded from spectral analysis utilising the high energy event rates of the respective detectors. 
All X-ray light curves are background subtracted, the background was taken from nearby source-free regions.
Spectral analysis of the EPIC data was carried out with XSPEC version 11.3 \citep{xspec} and is performed in the energy range 0.2\,--\,10.0\,keV. 
The RGS data of the individual observations is of very low SNR, we therefore merged the data from all three observations using the tool {\it rgscombine}.
We note that the X-ray source RX~J1556.4-3749 of unknown type lies in the read-out direction of the RGS, the
offset in the dispersed spectra is around -0.5\AA\,. Therefore RX~J1556.4-3749
could in principle contaminate the RGS spectra of RU~Lup, but being substantially absorbed as deduced from its EPIC spectra
its contribution to the here discussed spectral lines and especially the \ion{O}{vii} triplet is negligible. 
To reduce background and detector noise effects we extracted RGS spectra from the standard source region at phases of very low background (95\% PSF, background rate $<$\,0.25\,cts/s) 
as well as from the line core at low background (66\% PSF, background rate $<$\,1\,cts/s) and cross-checked the derived results. For
line fitting purposes we use the CORA program \citep{cora}, using identical line width and assuming Lorentzian line shapes.
Emitted line fluxes are corrected for absorption by using the {\it ismtau}-tool provided in the PINTofALE software \citep{poa}.

The data of the EPIC detectors are analysed simultaneously but not co-added for each observation and are modeled with a multi-temperature model 
assuming the emission spectrum of a collisionally-ionized, optically-thin gas
as calculated with the APEC code \citep{apec}. We find that a three-temperature model adequately describes the data.
Absorption in the circumstellar environment and in the interstellar medium is significant
and is applied in our modelling as a free parameter.
Abundances are modelled relative to solar photospheric values as given by \citet{grsa}. The low FIP (First Ionisation Potential) elements Al, Ca, Ni are tied to the Fe abundance;
for other elements with no significant features in the measured X-ray spectra solar abundances are used.
Since absorption, the coolest plasma temperature and abundances do not significantly differ between the observations, they are modelled with variable, 
but tied values to ensure a consistent analysis of the data.
We then derived the temperatures and volume
emission measures (EM=$\int n_{e}n_{H}dV$) of the individual plasma components for all three observation and calculated
corresponding X-ray luminosities from the resulting best fit models.
Some of the fit parameters are mutually dependent.  
This interdependence mainly affects the strength of absorption and emission measure of the cooler plasma at a few MK, but also emission measure components and abundances
of elements with emission lines in the respective temperature range.
Consequently, models with different absolute values of these parameters but only marginal differences in its statistical
quality may be applied to describe the data.

Our fit procedure is based on $\chi^2$ minimization, therefore spectra
were rebinned to satisfy the statistical demand of a minimum value of 15 counts per spectral bin.
All errors derived in spectral fitting are statistical errors given by their 90\% confidence range and were calculated separately
for abundances and temperatures by allowing variations of normalizations and respective model parameters.
Note that additional uncertainties may arise from errors in the atomic data and instrumental calibration which are not explicitly accounted for.

\section{Results}
\label{results}

\subsection{X-ray light curves and variability}

\begin{figure}[ht]
\includegraphics[width=90mm]{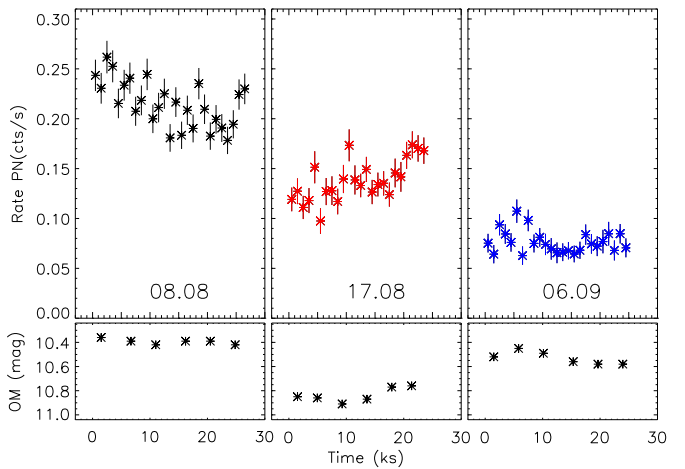}
\caption{\label{pnlc} Light curves of RU Lup determined with the PN instrument (1\,ks binning) in the energy band 0.2\,--\,5.0\,keV {\it (upper panels)} 
and OM UVW1 magnitude per exposure {\it( lower panels)}. }
\end{figure}

The {\it XMM-Newton} X-ray light curves of RU~Lup are shown in the upper panel of Fig.\,\ref{pnlc}. We observe a steady
decline in X-ray brightness by a factor of three over a month. Note that because of the
large time gaps between the observations (9 and 20 days respectively) this decline don't need to be monotonic. Indeed,
already in the first observation $L_{\rm X}$ declined by roughly 40\% within 25\,ks but started to rise again towards the end of the observation. 
During the second observation the count rate increased by roughly 50\% and short-term X-ray variability
is quite common in all observations.
We can rule out the presence of strong flaring within the individual observations,
however the August~08 light curve may be interpreted as a part of the decline phase of a larger flare. At any rate, the observed
overall decline during the {\it XMM-Newton} observations is very likely accidental due to sparse temporal sampling.

The OM UVW1 flux observed from RU Lup is plotted in the lower panel of Fig.\,\ref{pnlc}.  Note that derived magnitude
errors are in the range of 0.01\,mag and below the size of the shown symbols.  
We find variations in UVW1 brightness of up to 0.5\,mag between the three observations, however these variations do not correlate with
the observed X-ray brightness and have a far lower amplitude compared to the variations seen in the X-rays.
Additional UVW1 variability is observed within each observation of the order 0.1\,--\,0.15\,mag. Some of these variations appear to be 
correlated with the X-ray flux, possibly in the second and especially during the third and X-ray faintest observation. 
While a time-resolved X-ray spectral analysis
within each observation suffers from the poor signal and a correlation for six exposures might be a chance effect, the fact that the correlation is
most strongly present during the low-activity phases, where the soft X-ray component becomes more important, supports the view that at least some of the X-ray emission 
might be caused by the same physical process as the observed UV emission since the OM UVW1 is quite sensitive to flux emitted
from the accretion spot region. 

\subsection{Global spectral properties}

\begin{figure}[ht]
\includegraphics[height=87mm,angle=-90]{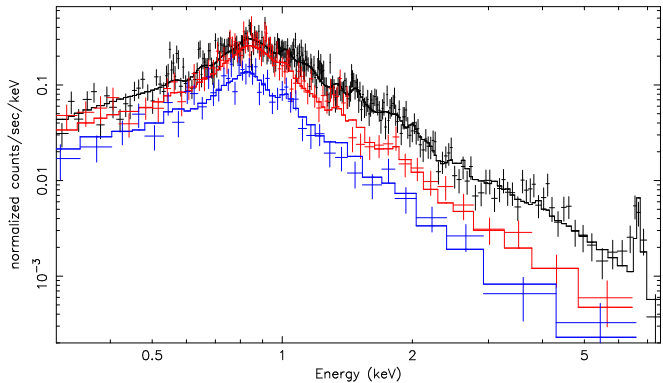}
\caption{\label{pnspec}X-ray spectra (PN) of RU Lup ({\it crosses}) and spectral fit ({\it histogram})
for the three observations; top down: 08/08 (black), 17/08 (red), 06/09 (blue).}
\end{figure}

To study the global spectral properties of the X-ray emission from
RU~Lup and investigate its changes we use the EPIC data.
In Fig.~\ref{pnspec} we show the PN spectra and the best fit models for the three observations to indicate data quality and spectral changes between the observations.
Obviously the spectrum is harder with increasing X-ray brightness and the plasma at hot temperatures is more prominent.
Very hot plasma ($\gtrsim 40$\,MK) is also evidenced by the presence of the 6.7\,keV iron line complex
in the data from August~8. In order to
quantify the spectral changes, we fit the EPIC data of the individual observations with a three-temperature model as described above 
and present the derived model parameters in Table\,\ref{specres}. 
Note that while absorption and abundances do not vary significantly between the exposures, fits of similar quality can be obtained with a cool component at very low
temperatures ($\sim$ 1\,MK) and a much larger EM, combined with moderately higher values for the oxygen abundance 
$A_{\rm O}$ and interstellar column density $N_{H}$.

We investigate in detail the decline in X-ray brightness throughout the observations and
find that there are only marginal changes in the EM of the cool component around 2\,MK; in contrast, the EM of the hot component decreases significantly by a factor of roughly six.
At its brightest phase the EMD is dominated by a hot plasma component at 30\,MK, and
its decline in EM is accompanied by a moderate cooling. In the third observation also the intermediate
plasma component is significantly weaker and slightly cooler compared to the other two observations. 
Consequently the average coronal temperature drops from 25\,MK over 15\,MK to 11\,MK during the campaign, mainly caused
by the decline of EM at temperatures around 20\,--\,35\,MK.
These findings suggest that the fainting of RU~Lup by a factor of three in emitted $L_{\rm X}$ is caused by a decline of its 
magnetic activity over a month, and hence that lack of any correlation between X-ray and UV brightness for the three observations is not surprising.

To compare our results with previous X-ray  measurements, we also calculate our model fluxes in the {\it Einstein} (0.5--4.0\,keV) and ROSAT (0.1--2.4\,keV) bands.  
We find that RU~Lup was actually in a phase of enhanced activity in autumn 2005, in agreement with the results from our spectral fitting.
Even during the third and X-ray darkest {\it XMM-Newton} observation, the determined flux of RU~Lup is nearly 20\% above the ROSAT value and similar to the {\it Einstein} upper limit.

From our spectral fits we derive a moderate absorption column of $N_{H}=1.8 \times 10^{21}$\,cm$^{-2}$ when compared to
typical values for CTTS \citep{rob07}. However, recalling that the optical/UV extinction measurements suggest columns of $\sim 10^{20}$\,cm$^{-2}$ and the fact 
that RU~Lup is viewed close to pole-on, the absorption value derived from the X-ray data appears extremely large. 
In order to reconcile X-ray and optical/UV absorption columns
requires an optically transparent medium with X-ray absorption, an issue we will return to in the next section.  

We find no strong overall metal depletion, however, some differences for individual abundances are present.
While the modelled abundances of medium FIP (O, Si, S) elements are moderately subsolar, the low FIP (Fe, Mg) elements show a higher abundance around or slightly above solar value 
and especially the high FIP elements (Ne) are enhanced. 
A high Ne abundance is commonly observed in CTTS and in active stars, but a high Fe abundance is rather atypical for a young CTTS. 
Depending on the specific model we find a Fe/O ratio of 1\,--\,1.5 and Ne/O of 2\,--\,2.5.
Recently, \citet{tel07} showed that a dependence on spectral type might be present, with G-type CTTS showing a higher Fe abundance 
(or Fe/O-ratio) as CTTS of spectral type K--M. RU~Lup's spectral type is usually classified as late K and its abundance ratios
are intermediate between their K--M-type and G-type population, in contrast to their findings.
We note that TW Hya, the prototypical accretion dominated CTTS and also of spectral type late K, shows likewise an Fe/O ratio of 1\,--\,1.5.
Thus spectral type may not be the only relevant parameter for abundance irregularities and e.g. accretion properties may have to be considered.

\begin{table}[!ht]
\setlength\tabcolsep{6pt}
\caption{\label{specres}Spectral fit results for RU~Lup, units are $N_{H}$ in $10^{21}$cm$^{-2}$, kT in\,keV, EM in $10^{52}$cm$^{-3}$
and $L_{\rm X}$ observed/emitted in $10^{29}$\,erg\,s$^{-1}$.}
\begin{center}
{
\begin{tabular}{lrrr}\hline\hline
Par. & 08/08 & 17/08 & 06/09\\\hline
$N_{H}$ & 1.8$^{+ 0.2}_{- 0.3}$ & 1.8$^{+ 0.2}_{- 0.3}$& 1.8$^{+ 0.2}_{- 0.3}$ \\[1mm]
Fe & 1.07$^{+ 0.13}_{- 0.20}$ &  1.07$^{+ 0.13}_{- 0.20}$ &  1.07$^{+ 0.13}_{- 0.20}$\\[1mm]
Si & 0.64$^{+ 0.29}_{- 0.27}$ & 0.64$^{+ 0.29}_{- 0.27}$ & 0.64$^{+ 0.29}_{- 0.27}$\\[1mm]
O &0.60$^{+ 0.19}_{- 0.27}$ & 0.60$^{+ 0.19}_{- 0.27}$  & 0.60$^{+ 0.19}_{- 0.27}$ \\[1mm]
Ne & 1.38$^{+ 0.35}_{- 0.31}$  &1.38$^{+ 0.35}_{- 0.31}$ & 1.38$^{+ 0.35}_{- 0.31}$\\[1mm]
kT1 & 0.18$^{+ 0.02}_{- 0.03}$ & 0.18$^{+ 0.03}_{- 0.03}$ & 0.18$^{+ 0.04}_{- 0.03}$  \\[1mm]
kT2 & 0.63$^{+ 0.02}_{- 0.02}$ & 0.63$^{+ 0.03}_{- 0.03}$ & 0.57$^{+ 0.03}_{- 0.08}$  \\[1mm]
kT3 & 2.97$^{+ 0.19}_{- 0.18}$ & 2.50$^{+ 0.42}_{- 0.40}$ & 2.31$^{+ 0.92}_{- 0.51}$  \\[1mm]
EM1 & 1.41$^{+ 0.72}_{- 0.36}$& 1.46$^{+ 0.63}_{- 0.50}$ &  1.55$^{+ 0.64}_{- 0.41}$ \\[1mm]
EM2 & 2.73$^{+ 0.18}_{- 0.16}$& 2.94$^{+ 0.27}_{- 0.20}$&  1.49$^{+ 0.13}_{- 0.42}$ \\[1mm]
EM3 & 8.72$^{+ 0.32}_{- 0.32}$ & 2.94$^{+ 0.36}_{- 0.33}$& 1.37$^{+ 0.31}_{- 0.18}$ \\ [1mm]\hline
$\chi^2_{red}${\tiny(d.o.f.)} & 1.03 (457) & 1.02 (147) & 0.90 (112) \\\hline
$L_{\rm X}$ & 13.2 (20.9)& 6.8 (12.5)& 3.4 (7.1)\\\hline
\end{tabular}
}
\end{center}
\end{table}

We also check for cool plasma that could be responsible for a possible UV/X-ray correlation within individual observations. 
Since moderate absorption is present in the X-ray spectra,
the required plasma temperatures have to be sufficiently high to be responsible for any significant change in the X-ray light curve.
Utilising the above derived value of $N_{H}$, we find that plasma
temperatures of roughly 2.5\,MK are sufficient, matching the temperature regime where accretion shock plasma is expected. 
In our 3-T models this corresponds to plasma mainly from the cool and partly from the medium temperature component.
This cooler plasma is obscured by hotter plasma during the more active and X-ray brighter first observation, but
becomes dominant especially in the third observation.
Therefore the plasma properties derived from the spectral fits are consistent with
the presence of a correlation between UV brightness and soft X-ray emission originating from accretion shocks. 
A stronger accretion component would result in a softening of the X-ray spectra, provided that the usually hotter coronal component is stable.
A softening of the X-ray emission with increasing brightness has been marginally detected during an {\it  XMM-Newton} observation of TW~Hya (\citet{rob06},
but the count rate of RU~Lup is not sufficient to calculate meaningful hardness ratios and investigate this possibility.

\subsection{Fluorescent emission}

\begin{figure}[ht]
\includegraphics[height=87mm,angle=-90]{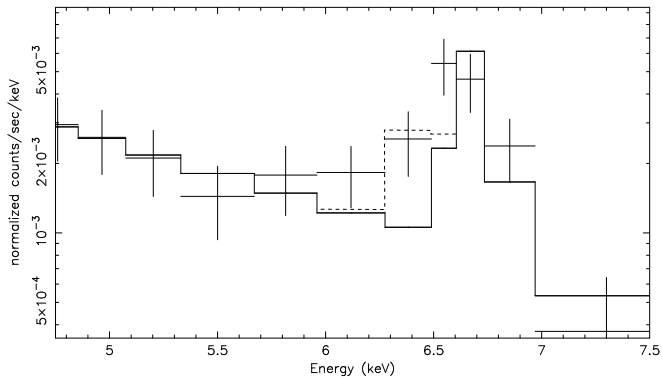}
\caption{\label{kaspec}
Spectral region covering the iron K$_{\alpha}$ line at 6.4\,keV and \ion{Fe}{xxv} line complex at 6.7\,keV, shown 
with the three temperature fit ({\it solid}) and with additional fluorescence line ({\it dashed}).}
\end{figure}

In the spectrum of the observation on August~08 some excess emission at energies below the 6.7\,keV iron line complex is visible,
which might be caused by fluorescence emission from iron K$_{\alpha}$. 
After inspection of the PN spectrum in the energy range 4.5\,--\,8.0\,keV we added a narrow (10\,eV) Gaussian component at 6.4\,keV to the three temperature model 
as shown in Fig.\,\ref{kaspec} and find that the
quality of the spectral fit to improves significantly from $\chi ^{2}_{red}$ = 1.33 (11~dof) to $\chi ^{2}_{red}$ = 1.03 (10~dof).
The derived flux from the improved spectral model in the fluorescence line is $5.5\pm4.3 \times 10^{-7}$~photons~cm$^{-2}$~s$^{-1}$
and the fluorescence line from iron K$_{\alpha}$ at 6.4\,keV is present at high significance ($>$\,90\%).

Under the assumption that the 
fluorescence emission from iron K$_{\alpha}$ is produced by the illumination of cooler material with X-ray photons
- in principle also electron excitation is possible - 
the exciting photons need to have energies above 7.11\,keV and the illuminated material must be at temperatures of $\lesssim$\,2MK.
While no large flare is present to produce these photons, the hot plasma component might be sufficiently bright to illuminate the star and the disk.
Additionally the pole-on geometry of RU~Lup is considered optimal for the production of fluorescence emission.
The required fluorescence efficiency is of at least 5\,\%, while for the best fit value up to 15\,--\,20\,\% efficiency are necessary.
These values are somewhat larger than the typical efficiencies 
of 3\,\% derived from Monte Carlo calculations \citep{bai79} to model iron fluorescence emission seen during solar flares. 
However, on the Sun only the photosphere is illuminated by the X-rays.
Larger efficiencies can be obtained from hidden X-ray sources, i.e. active regions on the far side of the star 
and/or with additional target material. A natural explanation would be the surrounding disk, which is likewise illuminated by the stellar X-ray emission.
A strong iron fluorescence line has also been detected in other young stellar objects like Elias29 \citep{fav05el} or in the several COUP sources in Orion \citep{tsu05}.
Theses observations
indicate the importance of X-ray emission for the ionisation and consequently temporal evolution of circumstellar disks.
While most previous detections are related to strong flaring, K$_{\alpha}$ emission from Elias29 was also observed in its quiescent state, similarly to the detection in RU~Lup.
Both sources share also the exceptional high coronal iron abundance, which has been interpreted by \citet{fav05el} as evidence for 
X-ray emitting plasma in magnetic flux tubes connecting the star and the circumstellar disk. Magnetic field lines between the star and its disk
are a natural ingredient of the magnetospheric accretion model and could provide a flow of fresh disk material into the corona.

\subsection{Oxygen lines - plasma density}

\begin{figure}[ht]
\begin{center}
\includegraphics[width=87mm]{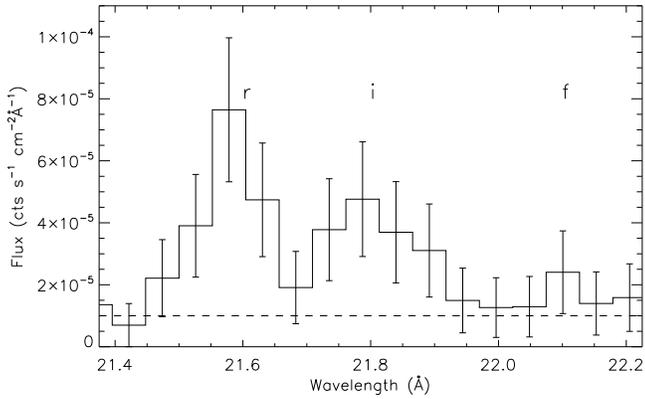}
\caption{\label{ovii}Fluxed \ion{O}{vii} triplet from RGS data (PSF core, rebinned), the background level is indicated by a dashed line.}
\end{center}
\end{figure}

We use high resolution RGS spectra to measure emission lines from \ion{O}{vii} and \ion{O}{viii},
i.e. the resonance\,(r), intercombination\,(i) and forbidden\,(f) lines in the He-like triplet 
of \ion{O}{vii} (21.6, 21.8, 22.1\,\AA) and the Ly\,$\alpha$ line of \ion{O}{viii} at 18.97\,\AA\,.
We search for cool excess plasma in CTTS compared to main-sequence stars via the \ion{O}{viii}/\ion{O}{vii}(r) line ratio, presented in Sect.\,\ref{ctts} and study the
density of the cool (2\,--\,3\,MK) plasma with the f\,/\,i\,-\,ratio in the He-like triplet of \ion{O}{vii}.
A density analysis of moderately hotter plasma at $\sim$~4\,MK could be carried out with the \ion{Ne}{ix} triplet,
but the unfortunately only moderate SNR of the RGS spectra and non negligible contamination from the second source (RX~J1556.4-3749) in this spectral range prevents this analysis. 
On the other hand, as shown in Fig\,\ref{ovii}, the \ion{O}{vii} triplet is clearly detected and shows an the intercombination line significantly stronger than the forbidden line.
We extract the line counts from the spectra obtained with the two extraction methods and list them in Table\,\ref{cora}. The derived 
physical properties are then calculated with line intensities corrected for effective area and
absorption as adopted from the EPIC model.
The plasma density is determined from the relation f\,/\,i$=R_{0}$\,/\,$(1+\phi/\phi_{c}+n_{e}/N_{c})$  with f and i being the respective line intensities,
$R_{0}=3.95$ is the low density limit of the line ratio,
$N_{c}$ the critical density and $\phi/\phi_{c}$ the radiation term, which is neglected in our calculations as discussed below.
The applied values are taken from \citep{pra81}.

\begin{table}[!ht]
\setlength\tabcolsep{4pt}
\caption{\label{cora}Measured line counts from RGS data for RU~Lup for the two extraction regions described in Sect.\,\ref{obsana}.}
\begin{center}{
\begin{tabular}{lrrrrr}\hline\hline
 & \multicolumn{2}{c}{OVIII} & \multicolumn{3}{c}{OVII - RGS1}\\
PSF & RGS1 &  RGS2 & r & i  & f\\\hline
Std. &28.7$\pm$8.1&33.0$\pm$8.1 & 22.7$\pm$7.3 & 20.1$\pm$6.9 & 8.3$\pm$5.6 \\
Core &23.8$\pm$6.4&26.3$\pm$6.2 & 22.9$\pm$6.1 & 15.3$\pm$4.0& 4.0$\pm$3.2 \\\hline
\end{tabular}}
\end{center}
\end{table}

We find a low f\,/\,i\,-\,ratio independent of the used PSF fraction, 
While actually the f-line is only marginally detected, formally 
the derived values are f\,/\,i\,$=0.26 \pm 0.23$ for the PSF core and 
f\,/\,i\,$=0.41 \pm 0.31$ for the standard region respectively. In both cases the \ion{O}{vii} f\,/\,i\,-\,ratio is substantially below unity,
resulting in a density of $n_{e}=4.4 (2.7) \times 10^{11}$cm$^{-3}$ for the two extraction methods.
The lower limit for the density is $\sim 2 (1.5) \times 10^{11}$cm$^{-3}$, on the other hand much higher
densities cannot be excluded due to the weakness of the forbidden line. 
The PSF core result is closer to the theoretically expected g-ratio (g=(f+i)/r) of around unity, but both methods agree within errors.
Thus RU~Lup is another CTTS that shows a high density in its cool ($\sim$~2\,MK) plasma component.
Two effects, namely the presence of an UV field and of coronal plasma at lower densities, may alter the actual density of the accretion shock plasma in either direction.
The strength of the UV field in the proximity of the X-ray emitting plasma is unknown, but its presence may influence the inferred plasma density.
A strong UV radiation field would lower the derived plasma densities, however strong UV radiation 
also has to be attributed to accretion shocks since the photosphere of a late-type star does not produce a sufficiently strong UV flux.
On the other hand, a coronal contribution to the
cool plasma, which is also present, results in an underprediction of the derived density for the X-ray emitting accretion shock plasma.
Therefore we conclude that the bulk of the cool X-ray emitting plasma in RU~Lup is generated in accretion shocks.

\section{An X-ray view on accretion and winds in CTTS}
\label{ctts}

RU~Lup is a high accretion rate CTTS and it is useful to compare its properties with those of other X-ray bright CTTS.  With the
increasing number of available medium and especially high resolution X-ray spectra for these objects we can study
signatures of two important processes in CTTS and in star-formation at large, accretion and outflows.

\subsection{Accretion, densities and cool excess}

While the X-ray emission of CTTS at higher energies (E~$\gtrsim$~1\,keV) is dominated by magnetic activity, density analysis with He-like triplets and other line diagnostics suggest that accretion
processes contribute significantly to the soft X-ray emission in many, and maybe even all CTTS. Since the disk truncation radii are usually large,
the accreting material is infalling with almost free fall velocities and hence post-shock temperatures of up to a few MK
are unavoidable. Such plasma is still relatively cool with respect to typical coronal temperatures of active stars and
should therefore lead to "cool excess" emission. This opens the possibility to investigate accretion scenarios via cool excess emission by comparing accreting stars
with stars where only coronal processes contribute to the X-ray emission. 

\begin{figure}[ht]
\includegraphics[width=85mm,height=50mm]{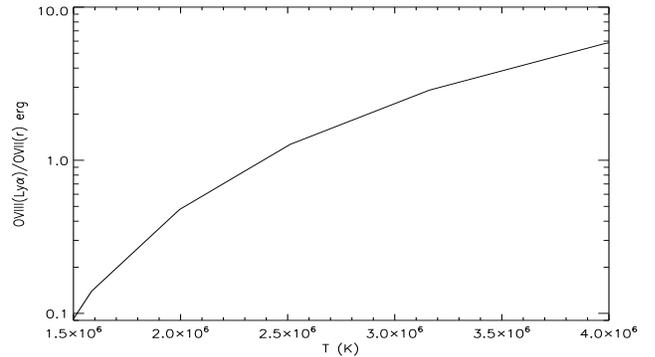}
\caption{\label{chianti} Theoretical \ion{O}{viii}(Ly$_{\alpha}$)/\ion{O}{vii}(r) line flux ratio calculated with the Chianti database vs. temperature.}
\end{figure}

Instead of using broad band fluxes we use the ratio of two strong lines that are commonly observed in stellar X-ray spectra,
the \ion{O}{viii} Ly$_{\alpha}$ line (18.97~\AA) and the \ion{O}{vii} resonance line (21.6~\AA) with peak formation temperatures of
$\sim$~3\,MK and $\sim$~2\,MK respectively.
In Fig.\,\ref{chianti} we show the \ion{O}{viii}(Ly$_{\alpha}$)/\ion{O}{vii}(r) line ratio as calculated with the CHIANTI V 5.2 code \citep{chi,anti}; 
it is very sensitive in the temperature range of interest and therefore a powerful diagnostic tool.
The emissivity curve of \ion{O}{vii}(r) dominates over the corresponding \ion{O}{viii}(Ly$_{\alpha}$)
curve at temperatures below $\sim$~2.5\,MK, and their ratio is therefore ideal to investigate cool excess plasma as expected from accretion shock models.

Specifically, we compare the ratios of CTTS and other accreting stars with sufficient signal in their high resolution X-ray spectra to those
measured in a large sample of main-sequence stars at various activity levels \citep{ness04}. Note that this sample contains several prominent non-accreting young stars like
AB~Dor, AT~Mic and AU~Mic. 
Results for the Herbig~Ae star AB~Aur are taken from
\citet{tel07}, the \ion{O}{vii}(r) line is inferred from their triplet flux assuming a g-ratio of unity, i.e.  (r=f+i). 
For the CTTS sample we adopt the values for BP~Tau, CR~Cha  and TW~Hya from
\citet{rob06} with a g-ratio of one for CR~Cha due to lower SNR, 
for MP~Mus from \citet{arg07}, for V~4046~Sgr from \citep{guenther06} with a distance of 80\,pc. T~Tau is re-analysed here in the same fashion as RU~Lup
(this work), where both extraction methods are shown.
All given values refer to the emitted line fluxes, i.e. they are corrected for the respective absorption column. 

\begin{figure}[ht]
\includegraphics[width=107mm]{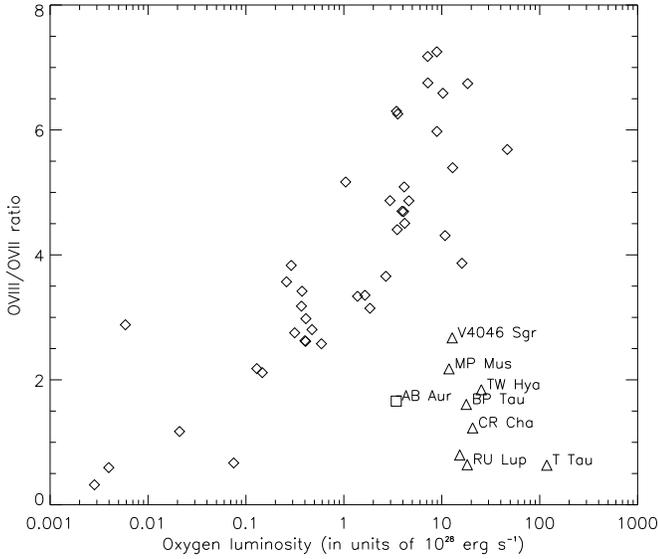}
\caption{\label{corr} Ratio of the emitted \ion{O}{viii}(Ly$_{\alpha}$)/\ion{O}{vii}(r) line flux vs. oxygen luminosity (\ion{O}{vii}(r)+\ion{O}{viii}(Ly$_{\alpha}$))
for main-sequence stars (diamonds), CTTS (triangles) and the Herbig Ae star AB~Aur (square).}
\end{figure}

In Fig.\,\ref{corr} we plot the absorption corrected \ion{O}{viii}(Ly$_{\alpha}$)/\ion{O}{vii}(r) line ratio vs. the total oxygen luminosity
(the power emitted in the  \ion{O}{viii}(Ly$_{\alpha}$) and \ion{O}{vii}(r) lines) for accreting stars and
main-sequence stars. The graph clearly shows 
that our sample CTTS and the main-sequence stars occupy quite different regions
and form two very well separated groups in the (\ion{O}{viii}/\ion{O}{vii}) vs. (\ion{O}{viii}+\ion{O}{vii}) diagram.
The correlation between the \ion{O}{viii}/\ion{O}{vii} line ratio and $L_{\rm oxy}$ for main-sequence stars is well known and 
caused by the on average higher coronal temperatures in the more active and X-ray brighter stars.

The important point now is that the inferred ratios of all accreting stars lie below the corresponding values for
main-sequence stars at the same oxygen luminosity. In other words, at a given oxygen luminosity an accreting star has a
smaller \ion{O}{viii}/\ion{O}{vii} line ratio than a non-accreting star. Since this ratio
depends only on temperature, there must be additional cool plasma radiating predominantly in the \ion{O}{vii} lines in accreting stars, i.e. an cool excess..
We emphasize that the above result is robust; investigating the \ion{O}{viii} and \ion{O}{vii} flux alone we find that the respective \ion{O}{viii} flux is 
compatible or slightly enhanced, whereas the \ion{O}{vii} flux is significantly enhanced in the accreting stars. 
While most sample CTTS have comparable oxygen luminosities
of $1-2 \times 10^{29}$\,erg\,s$^{-1}$, corresponding to roughly 5\,--\,10\% of their total $L_{\rm X}$ emitted in these two lines,
no correlation between cool excess and oxygen luminosity is present in our sample of accreting stars.
Thus the two groups show again different trends, suggesting a different origin of the observed emission.

In the magnetospheric accretion model a larger cool excess corresponds primarily to a higher accretion luminosity.
In addition the precise strength of the cool excess depends on the intrinsic temperature of the accretion shock plasma and the respective coronal contribution to the observed cool plasma, 
however both are affected by interdependencies and uncertainties that arise from modelling. 
Besides accretion other mechanisms have been proposed to produce an excess in the soft X-ray regime, 
for example disk and coronal stripping of the outer and hottest parts of the corona \citep{jar06},
disruption of hot plasma structures through accreted material \citep{aud05} or filling of coronal structures with cool material \citep{pre05}.
These mechanisms may likewise be present in young stellar objects, however, combining these finding with the often observed high densities in the cool plasma of CTTS,
an accretion shock leads to the required additional, persistent, cool, high density plasma component, in a completely natural way. 

Note that not all accreting stars generate significant X-ray emission in a high density environment via magnetically funnelled accretion streams.
While the exact density of the accretion plasma depends on the presence of UV-fields and the coronal contribution to the cool plasma,
all analysed CTTS with a moderate oxygen luminosity  exhibit cool plasma at high densities ($n_{e}\gtrsim 10^{11}$\,cm$^{-3}$).
This requires a strong funneling of the accretion stream, independent of the strength of their cool excess.
Contrary, the star with the largest cool excess and especially the largest oxygen luminosity, T~Tau itself, is the only T~Tauri star that exhibits cool plasma at a lower density and
actually stars like T~Tau are often not classified as CTTS, but as intermediate mass TTS (IMTTS) that are thought to be predecessors or analogs of Herbig AeBe stars.
In a related analysis of a TTS sample in Taurus-Auriga that shows a soft excess in accreting stars compared to WTTS presented by \citet{tel07}, 
the Herbig~Ae star AB~Aur is found as another example of an
accreting star that exhibits a cool excess and a low density cool plasma.
Both stars that show no high density plasma are of higher mass ($\gtrsim 2\,M_{\odot}$) than the sample CTTS ($\lesssim 1\,M_{\odot}$),
suggesting a scenario where accretion properties are linked to stellar properties, i.e. mass and radius.

\subsection{Accretion models vs. observations}
 
We now compare our results with expected accretion shock parameters from calculations based on the magnetically funnelled accretion model described in \citet{cal98}.
Discarding two object from the further discussion, V~4046~Sgr \citep[K5+K7, age\,$\sim$\,6\,Myr,][]{quast00} since it is a binary and the distribution of its X-ray emission 
among the binary components is unknown 
and CR~Cha (K2, age\,$\sim$\,1\,Myr) since its stellar parameters are 
only poorly constrained, we summarise the stellar properties of the accreting stars and their fitting results in Table\,\ref{star}. 
Filling factors as derived by \citet{cal98} are
in the range of $f=0.01$ for magnetically funneled accretion on CTTS, e.g. $f=0.007$ for BP~Tau, but can be up to an order of magnitude smaller or larger in individual stars.
Adopting their formulae and assuming free-fall velocity, i.e. the disk truncation radius to be much larger than the stellar radius,
the infall (post-shock  $n_{p}=4 \times n_{i}$) density is $n_{i}\sim \dot{M} \times f^{-1} \times (M^{-1/2}R^{-3/2})$. 
The accretion luminosity is $L_{Acc} \sim \dot{M} \times M/R$, the corresponding plasma temperature 
is $T\sim V^{2}\sim M/R$ and therefore assumed to be independent of funneling. 
Note that different shock temperatures also influence the observed  \ion{O}{viii}/\ion{O}{vii} ratio
and some uncertainties on the stellar parameters are also present for the investigated stars.
The coronal contribution to the X-ray emission from the cool plasma is approximated by the ratio EM1/EM2 in the respective 3-T models, in our sample stars the
corresponding temperatures are 2\,-\,3\,MK~(T1) and 6\,-\,7\,MK~(T2). Assuming similar shaped underlying stellar coronal EMDs as determined from active stars, 
this ratio traces the importance of the coronal contribution to the cool plasma. We denote EM1/EM2 ratios above 2 as 'weak', in the range 0.5--2 as 'moderate' and below 0.5 as 'strong'. 
Finally we roughly quantify the strength of the cool excess by comparing the \ion{O}{viii}(Ly$_{\alpha}$)/\ion{O}{vii}(r) line ratio for the coronal sources that exhibit the lowest 
value at a given oxygen luminosity and the respective accreting sources. 
We denote a cool excess as 'weak' for a ratio $\lesssim 2$, 'moderate' for a ratio of $\sim 2-4$ and 'strong' for larger ratios.

\begin{table} [ht!]
\setlength\tabcolsep{3pt}
\begin{center}
\caption{\label{star}Properties of accreting stars and their cool plasma.
Masses and radii from the references in this section and therein,
$M^{-1/2}R^{-3/2}$ in solar units, TS= $3.44\times  M/R$ (shock temp.), Cor.C.= coronal contribution.} 
\begin{tabular}{lrrrrrrr}
\hline\hline
Star &M & R & {\scriptsize M$^{-1/2}$R$^{-3/2}$}  &Density& TS& Cor.C. & Cool Ex. \\
 & $M_{\odot}$ &$R_{\odot}$ & & $10^{11}$cm$^{-3}$& MK & {\tiny (EM1/EM2)} & \\\hline
TW Hya & 0.7 & 1.0 & 1.20 &  vhigh 21. & 2.4 & weak(16.) &mod. \\
RU Lup & 0.8 & 1.7 & 0.50 &  high 4.4 & 1.6 & mod.(0.6)& strong \\
BP Tau  & 0.8 & 2.0 & 0.40 & high 3.2 & 1.4 & mod.(0.7) & mod.  \\
MP Mus  & 1.2 & 1.3 & 0.62 & high 5.0 & 3.2 & mod.(0.5) & weak \\
T Tau  & 2.4 & 3.6 & 0.09 & low 0.1 & 2.3 & weak(2.1) &strong \\
AB Aur  & 2.7 & 2.3 & 0.17 & low 0.1 & 4.0 & mod.(0.6) &weak\\\hline
\end{tabular}
\end{center}
\end{table}

Altogether the characteristics of the accreted plasma depend on an interplay of stellar properties and the accretion stream. The dependence on stellar properties
clearly favours the high density scenario for more compact low-mass stars, and predicts lower plasma temperatures and therefore larger soft excess for less contracted objects
as long as X-ray temperatures are reached in the accretion shock. Considering the accretion stream,
the ratio $\dot{M}/f$ is important for the resulting density, i.e.
large mass accretion rates and small filling factors are favoured in the higher density regime. A strong cool excess is produced by large mass accretion rates via
a large accretion luminosity. The accretion luminosity also increases for more compact objects, 
but when considering a cool excess as measured by \ion{O}{viii}/\ion{O}{vii} it conflicts with the corresponding higher shock temperatures.

There is a remarkable correlation between the cool plasma properties as expected in the magnetospheric accretion model from stellar parameters
alone and the observed properties for the six investigated stars. 
Especially the measured densities correlate well with the value of the quantity $M^{-1/2}R^{-3/2}$ 
suggesting that the ratio $\dot{M}/f$ does not vary extremely between the sample CTTS and that the cool plasma is indeed dominated by the accretion process in most cases.
The property $M^{-1/2}R^{-3/2}$ describes on the one hand the compactness of a star for a given mass but also some kind of general smallness.
The only star with very high density ($n_{e} >10^{12}$cm$^{-2}$), TW Hya, is also the only star with ({\tiny $M^{-1/2}R^{-3/2}$})\,$>$\,1, 
all stars with $ 0.4 \lesssim$\,({\tiny $M^{-1/2}R^{-3/2}$})\,$ \lesssim 0.6$ show high densities
($n_{e} >10^{11}$cm$^{-2}$) and the other two stars with low densities ($n_{e} \le 10^{10}$cm$^{-2}$) have ({\tiny $M^{-1/2}R^{-3/2}$})\,$<$\,0.2.
While within the high density stars the correlation is roughly linear,
the larger variations in density compared to those of $M^{-1/2}R^{-3/2}$ for the extreme cases of very high and low density suggest that an additional correlation
with $\dot{M}/f$ might amplify this trend. 
Consequently the funneling would have to be very effective for compact and tiny objects and much less pronounced for expanded or more massive stars.
Note, that the density of the cool plasma component does neither correlate with the shock temperature nor with the cool excess of the respective star.

The temperature of the accreted plasma is also determined by stellar mass and radius, but for the strength of the cool excess the coronal contribution and its X-ray brightness relative to the
one of the accretion spot is more important, since our definition of a cool excess involves \ion{O}{viii} emission which traces hotter plasma.
Necessarily the intrinsic temperature of the shocked plasma, which depends on the ratio M/R, has to be sufficiently low to produce a cool excess 
that is traced by enhanced \ion{O}{vii} emission.
The calculated shock temperatures are in the range between 1.5\,--\,4.0\,MK, i.e. the temperature range where the ionisation stage 
changes from a dominating \ion{O}{vii} contribution to a dominating \ion{O}{viii} contribution in a collisionally ionised plasma.
In two stars, MP~Mus and AB~Aur, the calculated shock temperature is already above 3\,MK and their cool excess is indeed weakest in our sample.
That the Herbig AeBe star AB~Aur shows some cool excess at all is surprising given its shock temperature of 4\,MK. 
This might be explained by inhomogeneous accretion spots that arise from recent modelling \citep{rom04}. These spots show
for example an intrinsic temperature distribution, an effect that might be more pronounced especially for less funneled accretion streams.
A strong cool excess is only observed in stars with shock temperatures below 2.5\,MK, but stars with similar shock temperatures may show quite different other accretion properties as
discussed exemplarily for the extreme cases T~Tau and TW~Hya. 
In T~Tau a large mass accretion rate of $3-6 \times 10^{-8}M_{\odot}$/yr \citep{cal04}, moderate shock temperatures
and a weak coronal contribution are present, producing the largest cool excess in our sample; its
lower plasma density requires a large filling factor and/or possibly involves other "cooling mechanisms" like mass loading of coronal structures as suggested by \citet{gue07}.
On the other hand, TW~Hya, the exceptional case of an accretion dominated CTTS shows no extreme cool excess. However, its shock temperature is only average and
in addition the lower mass accretion rate of $\sim 10^{-9}M_{\odot}$/yr \citep{muz00,ale02} contributes to TW~Hya's only moderate cool excess.
Its high plasma density then points to a strong funneling of the accreted plasma.
Further modifications for example through different disk truncation radii or levels of stellar activity seem to have only a minor impact on the cool excess at least in CTTS.
We note that CR~Cha and V~4046~Sgr on the whole fit into this picture since they are K-type low-mass stars and also show the expected high plasma density.
Admittedly our sample is not very large and completely lacking of any CTTS with very low mass ($\lesssim 0.5\,M_{\odot}$) due to the fact that they are usually X-ray darker since L$_{\rm X}$ 
correlates with mass in TTS. If also applicable in this mass regime, we expect early M-type CTTS to show rather high plasma densities that should be observable in X-rays,
if accretion rates are high enough and the accretion plasma is heated to sufficiently high temperatures.

One parameter, i.e. stellar age, is completely absent in the above discussion. 
The stellar evolution is of course "hidden" in the stellar radius for a given mass and consequently old ($\gtrsim$\,10\,My) and more compact CTTS like TW~Hya or MP~Mus
exhibit higher shock temperatures than young ($\lesssim$\,1\,My) CTTS like BP~Tau. 
Likewise the mass accretion rates of CTTS decrease in average throughout their evolution, further diminishing the cool excess with time.
While there is probably a large spread,
this trend is also present in the observations and none of the older CTTS, TW~Hya, MP~Mus and V~4046~Sgr, exhibit any strong cool excess. On the other hand, all older
CTTS show quite high plasma densities suggesting moderate mass accretion rates and a well funneled accretion stream.
This proposal of a unified scenario for the observed X-ray properties of accreting stars 
surely needs to be confronted with a larger sample to improve the statistics especially among the very low and intermediate mass regime, 
but provides a plausible explanation for the similarities as well as the differences in the observed characteristics of X-ray bright accreting stars.

\subsection{Outflows and anomalous X-ray absorption}

\begin{figure}[ht!]
\includegraphics[height=87mm,angle=-90]{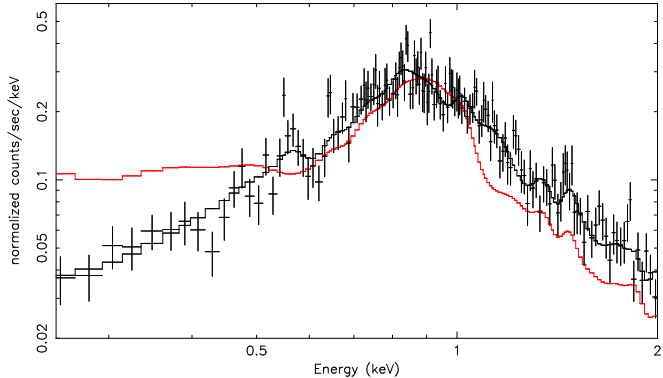}
\caption{\label{nh}X-ray spectra of the almost pole-on CTTS RU Lup. The histograms show our model (black) and the best fit model
with absorption set to values derived from the UV measurements (red) respectively,}
\end{figure}

In the X-ray spectra of the almost pole-on RU~Lup we measure an absorbing column density clearly incompatible with optical or UV measurements. 
In Fig.\,\ref{nh} we show the X-ray spectrum of RU Lup
as observed on August~08 with our best fit model and also with the best fit model using the
absorption value derived from UV measurements \citet{her05}. Column densities derived from optical observations
are usually comparable or even lower. Besides the fact that the applied model is quite unphysical since it does not contain any cool plasma, it still cannot explain 
the observed spectrum. Therefore we have to attribute the observed discrepancy to the presence of an unknown X-ray absorbing material along the line of sight.
The same effect was also detected in several other CTTS such as AA Tau \citep{schmitt07}, 
a CTTS viewed under an intermediate inclination of $\sim75\degr$ \citep{bouv99} and in the also pole-on CTTS TW~Hya \citep{rob06}, where the effect is an order of magnitude
lower and the discrepancy could be instead attributed to modelling or calibration uncertainties.

To reconcile the X-ray and UV/optical absorption observations requires an optically transparent, but sufficiently X-ray absorbing medium.  
This medium need to be free of any significant contribution of larger dust grains but has to contain a lot of X-ray absorbing material like gas.
Additionally  the absorption component has to be persistent and relatively steady over at least a month.
A good candidate for such a medium could be a strong outflow either from
the star, possibly powered by the accretion process \citep{matt05}, and/or from the inner disk \citep{koe00,ale05}.
Likewise the accretion streams may contribute to the additional absorption.
Another more exotic possibility could be a dust envelope with a very peculiar dust grain distribution containing sufficiently small grains. 
Considering the nearly pole-on view of RU~Lup, a wind scenario appears natural, in particular, since
strong, fast winds are well known for CTTS. However, the origin and geometry of the wind strongly depend on the underlying model
and detailed properties of CTTS winds are subject of considerable debate.

In general, a stellar wind requires lower mass outflow rates than a disk wind, since it originates from the stellar surface 
and therefore obscures the X-ray emitting regions more effectively. On the other hand, 
a stellar wind originating from coronal structures in analogy to the solar wind would have
temperatures of at least several 100,000\,K, much higher than the few 10,000\,K usually attributed to CTTS winds.
A hot wind has been claimed on the basis of asymmetric \ion{O}{vi} line profiles observed in TW~Hya by \citet{dup05}; however,
\citet{joh07} question this interpretation and argue for a much cooler wind in TW~Hya. 
Unfortunately our data does not allow to investigate the properties of the absorbing wind component in more detail.

If we adopt -- as the simplest absorption model -- that of a spherically symmetric wind with constant velocity, one derives a line-of-sight absorption column 
of $N_{H} = n_0 \times R_0$ with $n_0$ denoting the plasma density at the wind
base and $R_0$ its distance from the star.  Assuming the wind base to be essentially at the surface of the star and adopting the numbers of
RU~Lup, i.e. $N_H \sim 1.7 \times 10^{21}$\,cm$^{-2}$ and $R_0 = 1.7$ R$_{\odot}$, we find a base density of 
$1.4 \times 10^{10}$\,cm$^{-3}$. Further assuming this outflow to occur over 4$\pi$ steradian with a velocity of 300 km/s, results
in a mass loss rate of $\dot{M}_{wind} \sim 2 \times 10^{-9}$ M$_{\odot}$/yr.
However, the otherwise derived mass loss rates of CTTS are extremely uncertain, but are thought to be in the range 
0.01\,--\,0.1 of the respective mass accretion rates \citep{hart95,koe00}, which in turn are also
only poorly constrained and in addition time variable. Recent estimates for the mass accretion rates for RU~Lup are about
$5\pm 2\times10^{-8}$ M$_{\odot}$/yr \citep{her05}, 
and therefore a mass loss rate of $\dot{M}_{wind} \sim 2 \times 10^{-9}$ M$_{\odot}$/yr would perfectly fit the "expectations".  
Carrying out the same exercise for TW~Hya ($N_H \sim 3 \times 10^{20}$\,cm$^{-2}$) yields again consistent
results if one adopts mass accretion rates of $0.5-2\times10^{-9}M_{\odot}$/yr \citep{muz00,ale02}.
Note that outflow velocity, wind solid angle and location of the wind base appear in this model only as linear parameters and
moderate changes of their values do not change the rough estimations of the mass loss rates given above.
Accordingly, the lower X-ray absorption for TW~Hya would correspond to a more moderate wind, for RU~Lup both values
are roughly by a magnitude larger and may well be explained in a similar, scaled up scenario.
AA~Tau is viewed over the rim of a warped disk which partially occults the star at recurring phases \citep{bouv99} 
and associated material, accretion streams or outflows from the disk likely add significantly to its large and variable X-ray absorption \citep{schmitt07}.

We therefore conclude that the available X-ray data can indeed constrain the amount of material along the line of sight and
yield -- when adopting a simplified wind model with an outflow velocity of a few hundred km/s -- mass loss rates at the percent level
of the mass accretion rates. Interpreting therefore the observed X-ray absorption as arising from a wind thus yields
at least a physically consistent picture.

\section{Conclusions}
\label{sum}
From our studies of the X-ray emission from RU~Lup and other CTTS we draw the following conclusions:

\begin{enumerate}
\item RU~Lup is another example of a CTTS where cool high density plasma is present. 
The density of $3-4 \times 10^{11}$~cm$^{-3}$ as deduced from the \ion{O}{vii} triplet supports an accretion shock scenario
for the bulk of cool plasma. This cool plasma component is quite stable over a month.

\item Spectral variations indicate a change from an EMD that is clearly dominated by coronal activity in the first observation 
to a phase with a more equal distribution of cooler and hotter plasma in observation three. Many examples of coronal dominated CTTS have been detected so far, but RU~Lup
is a rare example where a change from state dominated by coronal activity to a state with a major contributing cool accretion component is observed over a month.
Its X-ray luminosity is not correlated with UV brightness
over the campaign, however during X-ray fainter phases there are indications for a correlation on timescales of several hours .

\item In all investigated accreting stars the characteristics of the cooler X-ray emitting plasma are influenced by the accretion process. 
An excess of cool plasma, as evidenced by a lower \ion{O}{viii}(Ly$_{\alpha}$)/\ion{O}{vii}(r) line-ratio, is present in our sample stars when compared to main-sequence stars.
The strength of the cool excess depends on an interplay of accretion shock luminosity and temperature as well as the omnipresent coronal contribution.
Cool, high density plasma is found so far exclusively in the low-mass CTTS sample $\lesssim 1 M_{\odot}$), 
while accreting stars with intermediate mass ($\gtrsim 2 M_{\odot}$) show lower densities.
Many aspects of the accretion process can be explained by stellar mass and radius and their evolution with time in a qualitative way.
We suspect a relation to mass accretion rates and especially amount of funneling, which produce the different properties of the accretion shock plasma 
that are seen in the respective X-ray spectra.

\item  We derive from X-ray spectra an absorption column of $1.8 \times 10^{21}$~cm$^{-2}$ for RU~Lup, 
roughly an order of magnitude above the value derive from Ly$_{\alpha}$ absorption.
To reconcile the optical extinction and X-ray absorption towards RU~Lup, one needs X-ray absorption in an optically almost transparent medium. 
Large discrepancies between absorption values derived from X-ray spectra and in the optical/UV wavelength regimes are also found in several other CTTS.
We suggest strong outflows/winds to be responsible, which originate either from the star and/or the disk. 
Clearly strong winds are present in RU~Lup and the nearly pole-on view does not favour anomalous dust material to be responsible for the excess absorption;
however, a contribution from the accretion flows is expected.
Using the required additional X-ray absorption columns and a simple wind model we derive mass loss rates of a few percent of the stellar mass accretion rates for CTTS. 
\end{enumerate}

\begin{acknowledgements}
This work is based on observations obtained with {\it XMM-Newton},
an ESA science mission with instruments and contributions directly
funded by ESA Member States and the USA (NASA).
J.R. acknowledges support from the DLR under grant 50OR0105. 
\end{acknowledgements}

\bibliographystyle{aa}
\bibliography{/data/hspc44/stch320/Pubs/jansbib}

\end{document}